\documentclass[cup5b]{caps}
\usepackage{psfig}
\usepackage{graphicx}
\usepackage{amssymb}
\usepackage{ociwsymp1e}
\HeadText{R. J. McLure and M. J. Jarvis}

\begin{document}

\pagenumbering{arabic}

\author[]{R. J. MCLURE$^1$ and M. J. JARVIS$^2$ \\
(1) Institute for Astronomy, Edinburgh University, Edinburgh EH9 3HJ, UK \\
(2) Sterrewacht Leiden, Postbus 9513, 2300 RA Leiden, The Netherlands}

\chapter{The Black Hole Masses of High-Redshift Quasars}

\begin{abstract}
A reliable method for estimating the black-hole masses of 
high-redshift quasars would provide crucial new information for 
understanding the nature and cosmological evolution of
quasars. In this proceedings we summarize the results of our 
recent paper (McLure \&
Jarvis 2002) which provides a virial black-hole mass estimator based
on rest-frame UV observables, thereby allowing reliable black-hole
mass estimates to be obtained out to redshifts of $z\sim2.5$ from
optical spectra alone. All cosmological calculations 
assume $\Omega_{m}=0.3, \Lambda=0.7, H_{0}=70$ kms$^{-1}$Mpc$^{-1}$.
\end{abstract}

\section{Introduction}
The underlying assumption behind the virial black-hole mass estimate 
is that the motion of the broad-line emitting material in AGN is 
virialized. Under this assumption the width
of the broad lines can be used to trace the Keplerian velocity 
of the broad-line gas, and thereby allow an estimate of the 
central black-hole mass via the formula
$M_{\rm BH}=G^{-1}R_{\rm BLR}V_{\rm BLR}^{2}$, 
where $R_{\rm BLR}$ is the broad-line region (BLR) radius and $V_{\rm BLR}$ is the
Keplerian velocity of the BLR gas. Currently, the most direct 
measurements of the central black-hole masses of powerful AGN 
are for 17 Seyferts and 
17 PG quasars for which reverberation mapping has provided a direct
measurement of $R_{\rm BLR}$; Wandel, Peterson, \& Malkan (1999) and Kaspi et 
al. (2000) respectively. 

An important result of these studies is the
discovery of a correlation between $R_{\rm BLR}$ and the monochromatic 
AGN continuum luminosity at 5100\AA\, 
(eg. $R_{\rm BLR}\propto \lambda L_{5100}^{0.7}$; Kaspi et al. 2000). 
By combining this luminosity based $R_{\rm BLR}$ estimate with a measure
of the BLR velocity based on the FWHM of the H$\beta$ emission line, 
it is now possible to produce a virial black-hole mass estimate from a 
single spectrum covering H$\beta$. This technique 
has recently been widely employed to investigate how the
masses of quasar black holes relate to the
properties of the surrounding host galaxies (eg. McLure \& 
Dunlop 2001, 2002; Laor 2001) and the radio luminosity of the 
central engine (Dunlop et al. 2003; Lacy et al. 2001). 

However, because the H$\beta$ emission line is redshifted into the
near-infrared at a redshift of $z\sim1$, it is observationally
expensive to use H$\beta$ to estimate the black-hole masses of $z>1$
quasars, with the vast majority of current studies 
concentrating on samples at $z\leq0.3$. In this proceedings we 
summarize the evidence presented in McLure \& Jarvis (2002) which 
suggests that the combination of 3000\AA\, continuum luminosity and Mg~II
FWHM can provide a reliable UV virial mass estimator in the redshift 
range $0.3<z<2.5$ from straightforward optical spectroscopy.

\begin{figure}
\psfig{file=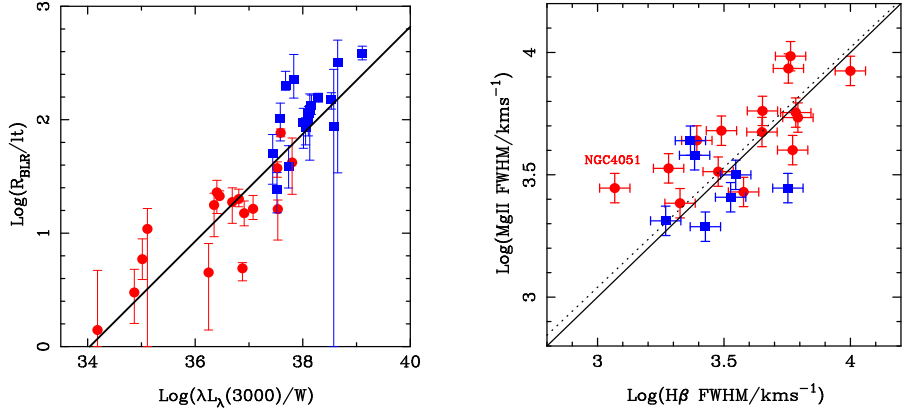,width=12cm,angle=0}                   
\caption{The correlation between broad-line radius and AGN continuum
luminosity at 3000\AA\,. The 17 PG quasars are shown as squares and 
the 17 Seyfert galaxies are shown as circles.
The best-fitting BCES bisector fit is shown as the solid line, and 
corresponds to $R_{\rm BLR}\propto \lambda L_{3000}^{0.47}$. Fig. 1.1b.
A Log-Log plot of Mg~II FWHM versus H$\beta$ FWHM for the 22 objects 
from the RM sample for which it was possible to obtain measurements of both
line-widths. The solid line is an exact 1:1 relation. The dotted line is
the BCES bisector fit (Akritas \& Bershady 1996), excluding NGC 4051,
and has slope of
$1.02\pm0.14$. NGC 4051 is a well known example of a narrow-line
Seyfert galaxy.}
\end{figure}

\section{The Radius-Luminosity Relation}
Fig. 1.1a. shows our analysis of the $R_{\rm BLR}-\lambda
L_{\lambda}$ relation at 3000\AA\, using a combined 
sample of 34 quasars and Seyfert galaxies with reverberation mapping 
measurements of $R_{\rm BLR}$ (RM sample, Kapsi et al. 2000; Wandel et al.
1999). The UV luminosities of the quasars are calculated from the 
data of Neugebauer et al. (1987), while the UV luminosities for the
Seyfert galaxies are derived from our analysis of archival IUE
spectra. The bisector fit is shown as the solid line in Fig. 1.1a, and is
equivalent to:
\begin{displaymath}
R_{\rm BLR}=(25.2\pm3.0) \left[\lambda L_{3000}/10^{37}W
\right]^{(0.47\pm0.05)}
\end{displaymath}
\noindent
where $R_{\rm BLR}$ is in units of light-days. We note here that the slope
of the $R_{\rm BLR}-\lambda L_{3000}$ relation ($0.47\pm0.05$) is entirely
consistent with that expected for a constant
ionization parameter (0.5). Furthermore, the scatter around the
$R_{\rm BLR}-\lambda L_{3000}$ relation is smaller than that associated
with the established relation based on 5100\AA\, luminosity (see McLure
\& Jarvis (2002) for details).
\section{Estimating the BLR Velocity}
The main reason for adopting Mg~II as the UV tracer of BLR velocity is
that, like H$\beta$, Mg{\sc ii} is a low-ionization line. Furthermore,
due to the similarity of their ionization potentials, it is reasonable 
to expect that the Mg{\sc ii} and H$\beta$ emission lines are 
produced by gas at virtually the same radius from the central
ionizing source. This assumption is directly tested in Fig. 1.1b. which
shows a plot of Mg~II FWHM versus H$\beta$ FWHM for the 22 objects with
reverberation mapping results (RM sample) for which it was possible to
obtain Mg~II FWHM measurements. Fig. 1.1b confirms that the FWHM of Mg~II
and H$\beta$ do follow a 1:1 relation. 
This has two important consequences. Firstly, it
allows us to directly adopt the $R_{\rm BLR}$ estimates from the
correlation between $R_{\rm BLR}$ and 3000\AA\, continuum
luminosity. Secondly, because the
line-widths of Mg~II and H$\beta$ should trace the same BLR velocities, we are
able to simply substitute the FWHM of Mg~II for that of H$\beta$ in the
virial mass estimator
\section{The UV Black Hole Mass Estimator}
\begin{figure}
\psfig{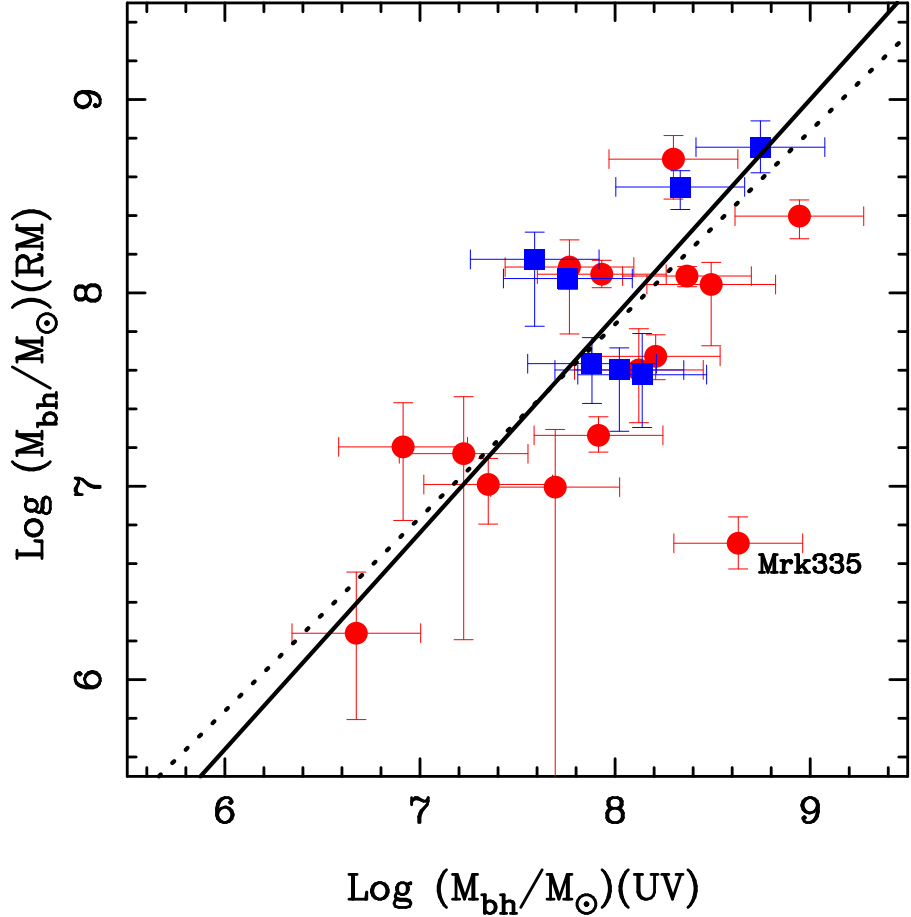}     
\caption{Full reverberation mapping black-hole mass
estimate (based on $R_{\rm BLR}$ measurements and the rms H$\beta$ FWHM)
plotted against the UV mass estimate (based on the $R_{\rm BLR}-\lambda
L_{3000}$ relation of Fig. 1.1a. and the FWHM of Mg{\sc ii}). 
The solid line is the BCES bisector fit, excluding
Mrk 335 a narrow-line Seyfert, and has a slope of $1.12\pm0.22$. 
The dotted line is the adopted linear relation. Symbols as in Fig. 1.1.}
\end{figure}

Having determined the $R_{\rm BLR}-\lambda
L_{3000}$ relation and the 1:1 scaling between H$\beta$ FWHM and
Mg~II FWHM, we are now in a position to derive our UV black-hole mass estimator.
In Fig. 1.2 we show the reverberation
mapping black-hole mass estimate versus the new UV mass estimate, which uses 
the $R_{\rm BLR}-\lambda L_{3000}$ relation to estimate
$R_{\rm BLR}$ and the Mg{\sc ii} FWHM to trace the BLR velocity. 
The solid line in Fig. 1.2 is the bisector fit to the data,
excluding Mrk 335, and has the form: $
\log M_{\rm BH}(RM)\propto1.12(\pm0.22)\log M_{\rm BH}(UV)$
, consistent with a linear relation. In light of this, the dashed line
in Fig. 1.2 shows the best-fitting linear relation which is adopted as
our final calibration of the UV virial mass estimator. 
In terms of a useful formula the final calibration of the UV 
black-hole mass estimator is therefore:
 
\begin{displaymath}
\frac{ M_{\rm BH}} {\mbox{\,$\rm M_{\odot}$}}  =3.37\left(\frac{\lambda
L_{3000}}{10^{37}{\rm W}}\right)^{0.47}\left(\frac{FWHM(Mg~II)}
{{\rm kms}^{-1}}\right)^{2}
\end{displaymath}
\noindent
Excluding Mrk 335, 
the mean difference between the reverberation and the UV estimator is :
$<\log(M_{\rm BH})(RM)-\log(M_{\rm BH})(UV)>=0.00\pm0.40$ ($1\sigma$). 
Provided the RM sample is representative of broad-line AGN, we conclude that 
the UV black-hole estimator can reproduce the reverberation 
black-hole mass to within a factor of 2.5 $(1\sigma)$. 
\section{Application to the MQS and LBQS}
\begin{figure}
\psfig{file=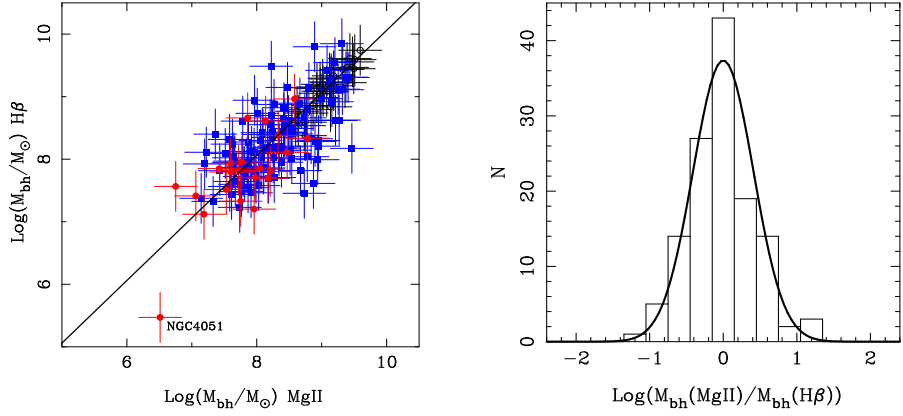,width=12cm,angle=0}     
\caption{The optical (H$\beta$) versus UV (Mg{\sc ii}) virial 
black-hole estimators for 150 objects from the RM (filled circles), 
LBQS (filled squares) 
and MQS (open circles) samples. The solid line is the
BCES bisector fit to the 128 objects from the MQS and LBQS samples and
has a slope of $1.00\pm0.08$. The outlying NLS NGC4051 has been 
highlighted. Fig. 1.3b Histogram of $\log M_{\rm BH}(Mg~II)-\log M_{\rm BH}(H\beta)$ for 
the 128 objects from the LBQS and MQS shown in Fig. 1.3a. Also shown is the 
best-fitting gaussian which has $\sigma=0.41$.}
\end{figure}

In Fig. 1.3. we show the optical black-hole mass estimator plotted
against the new UV black-hole mass estimator for a combined sample of
150 objects, comprising 99 from the Large Bright Quasar Survey (Forster et al. 2001), 29 from the radio-selected Molonglo Quasar Sample (Baker et al. 1999) 
and the 22 objects of the RM sample. Also shown is the BCES 
bisector fit to the LBQS and MQS objects which has the form: 
\begin{displaymath}
\log M_{\rm BH}(H\beta)=1.00(\pm0.08)\log M_{\rm BH}(Mg~II)+0.06(\pm0.67)
\end{displaymath}
\noindent
which, as expected, is perfectly consistent with a linear relation. In
Fig. 1.3b we show a histogram of 
$\log M_{\rm BH}(Mg~II)-\log M_{\rm BH}(H\beta)$
for the 128 objects from the LBQS and MQS. The solid line shows the
best-fitting gaussian which has $\sigma=0.41$. These results 
lead us to conclude that, 
compared to the traditional optical black-hole mass estimator, the 
new UV estimator provides results which are unbiased and of equal
accuracy.
\section{Conclusions}
The main conclusions of this study can be summarized as follows:

\begin{itemize}

\item{The correlation between $R_{\rm BLR}$ and 3000\AA\, continuum
luminosity is found to display less scatter than the established
correlation with 5100\AA\, luminosity, and to be consistent with the 
$R_{\rm BLR}\propto \lambda L_{\lambda}^{0.5}$ relation expected for a 
constant ionization parameter.}

\item{Combining the $R_{\rm BLR}-\lambda L_{3000}^{0.47}$ relation with the
FWHM of Mg~II produces a virial black-hole mass estimator based on
rest-frame UV observables which is capable of reproducing black-hole
masses determined from reverberation mapping to 
within a factor of 2.5 ($1\sigma$)}

\item{An application to objects from the LBQS \& MQS demonstrates 
that the new UV black-hole mass estimator produces results
which are unbiased, and of equal accuracy to the established 
optical (H$\beta$) black-hole mass estimator.}
\end{itemize}

\end{document}